# Injectable Spontaneous Generation of Tremendous Self-Fueled Liquid Metal Droplet Motors in a Moment

You-You Yao [a)*], Lei Sheng [a)*] and Jing Liu [a,b)**]

[a)] Department of Biomedical Engineering, School of Medicine,
Tsinghua University, Beijing 100084, China

[b)] Key Laboratory of Cryogenics, Technical Institute of Physics and Chemistry,
Chinese Academy of Sciences, Beijing 100190, China

* These two authors contribute equally to this work.

**Address for correspondence:
Dr. Jing Liu
Department of Biomedical Engineering,
School of Medicine,
Tsinghua University,
Beijing 100084, P. R. China
E-mail address: jliubme@tsinghua.edu.cn
Tel. +86-10-62794896
Fax: +86-10-82543767

**Abstract:**

Micro motors that could run in liquid environment is very important for a variety of practices such as serving as pipeline robot, soft machine, drug delivery, or microfluidics system etc. However, fabrication of such tiny motors is generally rather time and cost consumptive and has been a tough issue due to involve too many complicated procedures and tools. Here, we show a straightforward injectable way for spontaneously generating autonomously running soft motors in large quantity. A basic fabrication strategy thus enabled is established and illustrated. It was found that, injecting the GaIn alloy pre-fueled with aluminum into electrolyte would automatically split in seconds into tremendous droplet motors swiftly running here and there. The driving force originated from the galvanic cell reaction among alloy, aluminum and surrounding electrolyte which offers interior electricity and hydrogen gas as motion power. This finding opens the possibility to develop injectable tiny-robots, droplet machines or microfluidic elements. It also raised important scientific issues regarding characterizing the complicated fluid mechanics stimulated by the quick running of the soft metal droplet and the gases it generated during the traveling.

**Keywords:** Droplet motor; Fluid injection; Liquid metal; self-fueled soft matter; Tiny machine; Injectable device.

## 1. Introduction

The room temperature liquid metal (RTLM) is a kind of interesting material owning extraordinary chemical or physical properties that had not been fully explored before. With the increasing understanding of such material, RTLM has gradually been applied in a variety of newly emerging areas such as energy harvesting [1], chip cooling [2], microfluidics [3,4], and printed electronics [5] etc. Among the many issues ever tackled, the RTLM droplet is especially an interesting topic worth of pursuing [6-7]. And important endeavors have been made on the preparation of such metal sphere. Through exquisite control of the flow focusing, fluid velocity, and viscosity etc., various droplets with multiple sizes and shapes can be fabricated via microfluidic system [6]. Meanwhile, researchers also developed channelless way to directly form liquid metal spheres through injecting a liquid metal stream into solution with surfactant [7].

One critical role a liquid metal sphere could play is to make soft motor. As it was realized, micro-motors are offering great potential for many newly emerging areas [8-11]. Several important propulsion mechanisms were disclosed as catalytically generated gas bubbles or external electromagnetic field [12-14]. Recently, it was found that the liquid metal sphere in the electrolyte can be controlled to rotate, move and change shape via external electric field [15-17]. A breakthrough invention thus enabled is the liquid metal pump [17] which works just like a motor. So far, conventional ways for making micro motors is still somewhat complex due to request too many complicated procedures and equipment. While electricity actuated rotation of liquid metal sphere may serve as a motor, it is still not easy to precisely control the electrical field as desired especially when the motor size is rather small.

During our large number of experiments on testing, processing and manipulating the liquid metal fluid, we occasionally found an unusual phenomenon that injecting the RTLM pre-fueled by aluminum into the electrolyte would quickly make tremendous liquid metal tiny-motors. This letter is dedicated to present the most typical findings towards quickly fabricating tiny droplet motors through fluid injection and the mechanisms lying behind.

## 2. Materials and Methods

To fabricate the liquid metal tiny-motors, we first prepared eutectic GaIn (EGaIn) alloy with 74.5%Ga, 25.5%In in weight. Although different RTLM alloy and matching metals can be adopted as the candidate materials, only EGaIn and aluminum will be focused here for brief. According to former work [18], aluminum can be easily corroded by the gallium or its alloy. Therefore, we added 0.27g aluminum piece to 3mL EGaIn to make functional solution. To guarantee the fabrication quality, the whole process was performed in the 0.2 mol/L NaOH solution which can help remove the oxidation film on both aluminum and EGaIn and thus enhance their fusion. Then through the mixture, an alloy of Al-EGaIn alloy with different configurations can be formed (Fig. 1a). In principle, there may form different mixtures for the final functional solution such as total, partial, discrete or zero immersing of aluminum into EGaIn, as illustrated in Fig. 1b. This would affect the specific running of the tiny motors thus generated. For the case that no Al-electrolyte contact exists, the electric double layer on liquid metal droplet would remain stable. And no movement will be induced. While for the case the fuel is exposed to the electrolyte, the electric double layer of the RTLM droplets would become unbalanced and motor locomotion will be induced. In this case, we called such alloy as pre-activated functional solution. It was then sucked into a syringe for later use (Fig. 1c).

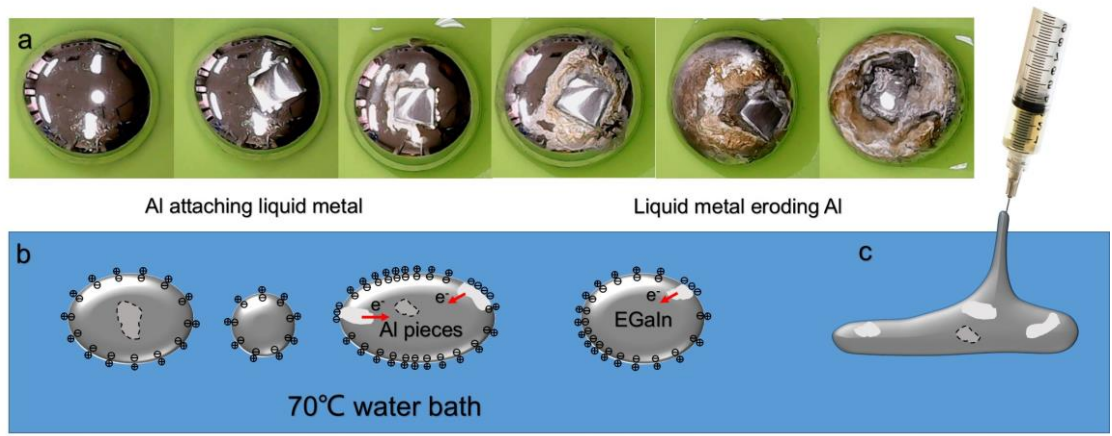

FIG. 1. Preparation of pre-activated liquid metal solutions. (a) Eroding fuel aluminum pieces into EGaIn alloy. (b) Electrical double layer charges distribution on the surface of liquid metal sphere immersed in the electrolyte. (c) Suction of liquid metal functional solution into a syringe for subsequent injection fabrication.

## 3. Results

The above functional solution can be injected into 0.2mol/L NaOH alkaline solution to make tiny motors. Through pushing the syringe piston forward, the liquid metal would jet into the electrolyte and breakup into a large number of droplets (Fig. 2a) in 1 or 2 seconds. Such droplet formation mechanism is due to the high surface tension of the liquid metal stream, as revealed before [7]. Interesting enough, a surprising phenomenon that had never been observed before is that such droplets just keep running here and there, representing tremendous tiny motors. According to specific needs, various quantities of liquid metal motors can be generated. With designed needle apertures, one can obtain either smaller or bigger tiny-motors. Here, only standard syringe aperture diameter around 0.3mm is focused for brief. To evaluate the basic running modalities of the tiny motors, we designed and machined a PMMA material based O-shaped open-top channel with width of 8mm (Fig. 2b) which is to contain the solution inside and to restrain the locomotion of the liquid metal motors. The alkaline solution was maintained at constant temperature by 70 ℃ water bath so as to guarantee an appropriate chemical reaction rate. Because of the low thermal conductivity 0.1884W/m·℃ of PMMA, there is an evident temperature gradient between the water both and the channel solution with later around 38 ℃. The liquid metal motor was found to work well above certain temperature such as at this specific value. Except for investigating the liquid metal tiny motors in large quantity (Fig. 2a), we also made two individual droplet motors (Fig. 2b) for detailed analysis purpose. The dynamic motion images were recorded by digital camera.

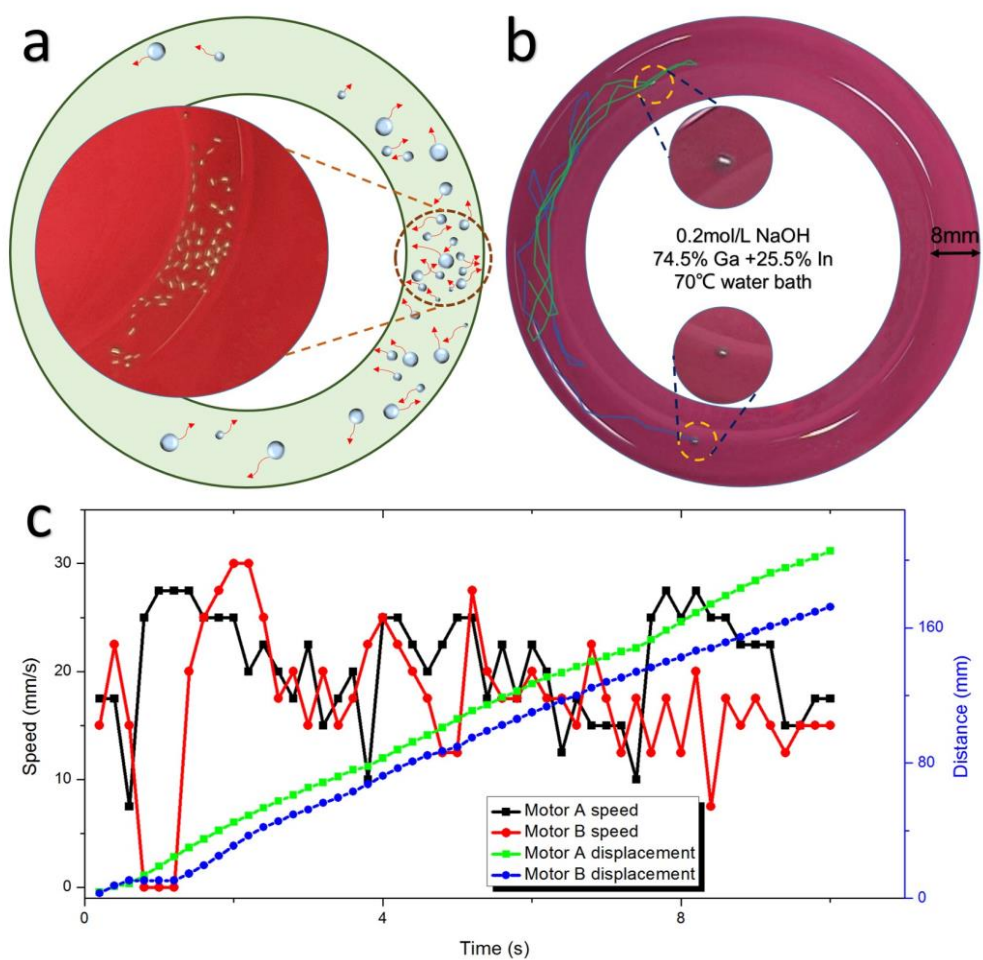

FIG. 2. Characterization of the locomotion of typical liquid metal tiny motors. (a) Diagram of many motors running in PMMA open top channel. (b) Motion trajectory of two individual motors. (c) Instantaneous velocities and displacements of two individual motors.

According to the experiments, there exist certain differences between the locomotion characteristics of a cluster of motors (Fig. 2a) and that of two separate droplet motors (Fig. 2b). When a cluster of droplets exercise in a closed space, they tend to crowd and affect each other.

The locomotion of each droplet then becomes rather hard to discern. While a single sphere or two could distinctly display their running trajectory. In this sense, Fig. 2b was used just to characterize the basic behaviors of the tiny motors such as locomotion path, running speed and lasting time etc. Here, the relative motion velocity was defined as distance that a motor travels in one second.

The diameters of the two RTLM spheres in Fig. 2b are measured as 1 millimeter. Through image and data processing, we could track the trajectory of two droplets Motor A and B and calculate their transient velocities. As it was quantified (Fig. 2c), the droplet motors swim quickly in aqueous solution in the open-top channel with a pretty high but somewhat random velocity on the order of tens of mini-meters per second. Such non-uniform velocity is caused by the dynamic position of aluminum fuel in the EGaIn droplet. Overall, the magnitude for the average velocity of the motor reached a very high magnitude 10-30mm/s (Fig. 2c), which is somewhat an unusual moving capability. Further, because of the high surface tension of the liquid metal EGaIn, the tiny motors are mostly shrunken into spheres with a diameter from less than half millimeter to millimeters. For a single motor, its lifetime can last for more than a half hour. Depends on different alloy combinations, such lifetime is regulatable.

Compared with a cluster of tiny motors (Fig. 2a), a single motor has a somewhat unstable motion trajectory which appears as inflexions, cuspidal points or retrace. The motor may run like shaking and vibrating with a straight velocity component. Sometimes it may hold standstill for a while, and then re-activate again suddenly. The typical delay time is approximately 0.5 second or less. Analogous phenomenon was also found that when injecting a large amount of droplets into the container, most tiny particles could keep standstill for a very short period of time and then resume to alive full of energy. Overall, motors in a cluster tend to be easily influenced by other motors running nearby.

For quantifying a cluster of droplet motors, image processing is performed to clarify how droplets aggregate and scatter in different positions (Fig. 3). With too many tiny motors assembling together, it is rather hard to identify the traveling path and velocity for each of the specific droplet motors. As an alternative, such original snapshots (see upper part image at left hand side of Fig. 3) were converted into gray-scale maps, utilizing the gray level difference between luminous spheres and the bleak background. The gray-scale maps are ulteriorly transformed to black and white images (see lower part image at left hand side of Fig. 3). The images are divided into four quadrants which have the same location as Cartesian coordinate system. By counting the white pixel numbers in each quadrant, we can estimate the droplet motors' transient distribution. It is interesting to find that, the transient pixel numbers in each of the quadrant regions I, II, III and IV are different and appear as an oscillating curve. When two clusters meet, their movements are disturbed and they may change their advancing direction or just run into opposite direction. Such process takes an approximate 8 seconds period (Fig. 3). This indicates that the liquid metal motors are not uniformly distributed throughout the whole channel. And the metal cluster may choose to stay more in certain quadrant.

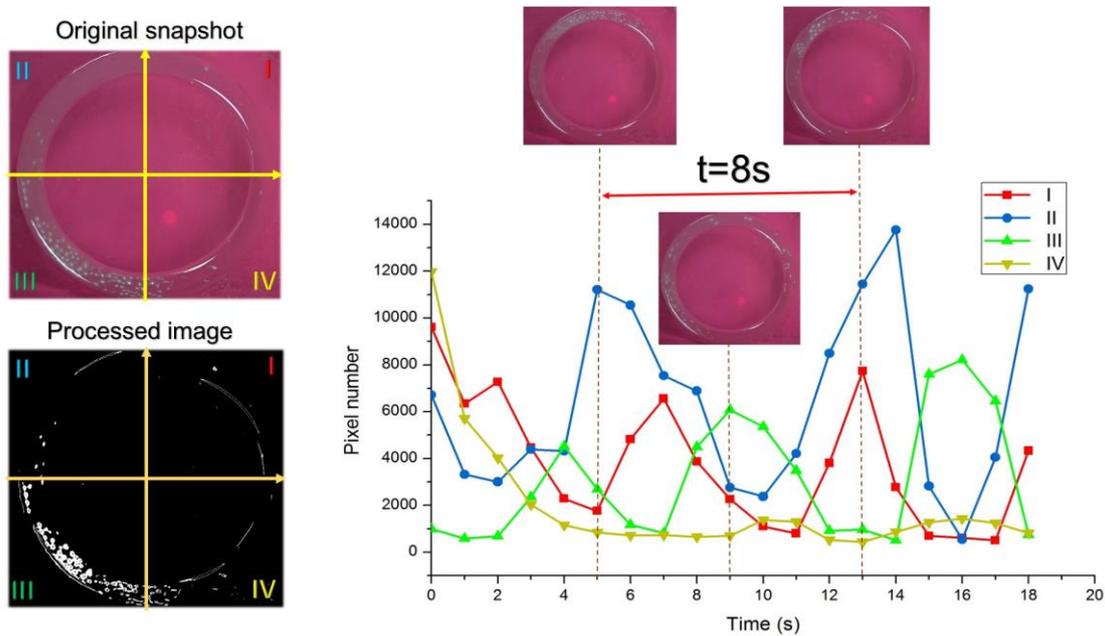

FIG. 3. Locomotion of a cluster of liquid metal tiny motors in the open-top channel. Here, each image processed as black and white is placed below its original image. Note: The channel width is fixed as 8mm throughout the whole experiments.

Presented in Fig. 4 is a sequential motion of the liquid metal motors in large quantity. Here, the time interval for recording the image is one second and the total record lasts about 20 second. These images are analyzed on Matlab calculation platform. From these subsequent figures, one can observe that, initially (t=0), when the liquid metal pre-fueled with aluminum was injected into the electrolyte, it would automatically form a large amount of droplet spheres. Then the droplets start to run along the channel (see picture at t=1s). One can see a stream of droplet motors crowding together while keep running (t=2s, 3s, 4s). They may run clockwise (t=3s, 4s) and then anticlockwise (t=5s, 6s, 7s). Such motor clusters may also depart themselves and run oppositely (t=7s, 8s, 9s). Depending on the specific injection, the droplets may not necessarily be in the same size. Through measuring the droplet diameters, one can find that the droplet motors at different location take different average diameters such as from D=0.54mm, 0.84mm to 1.29mm (Fig. 4) etc. The variation comes from which part of the droplets was sampled.

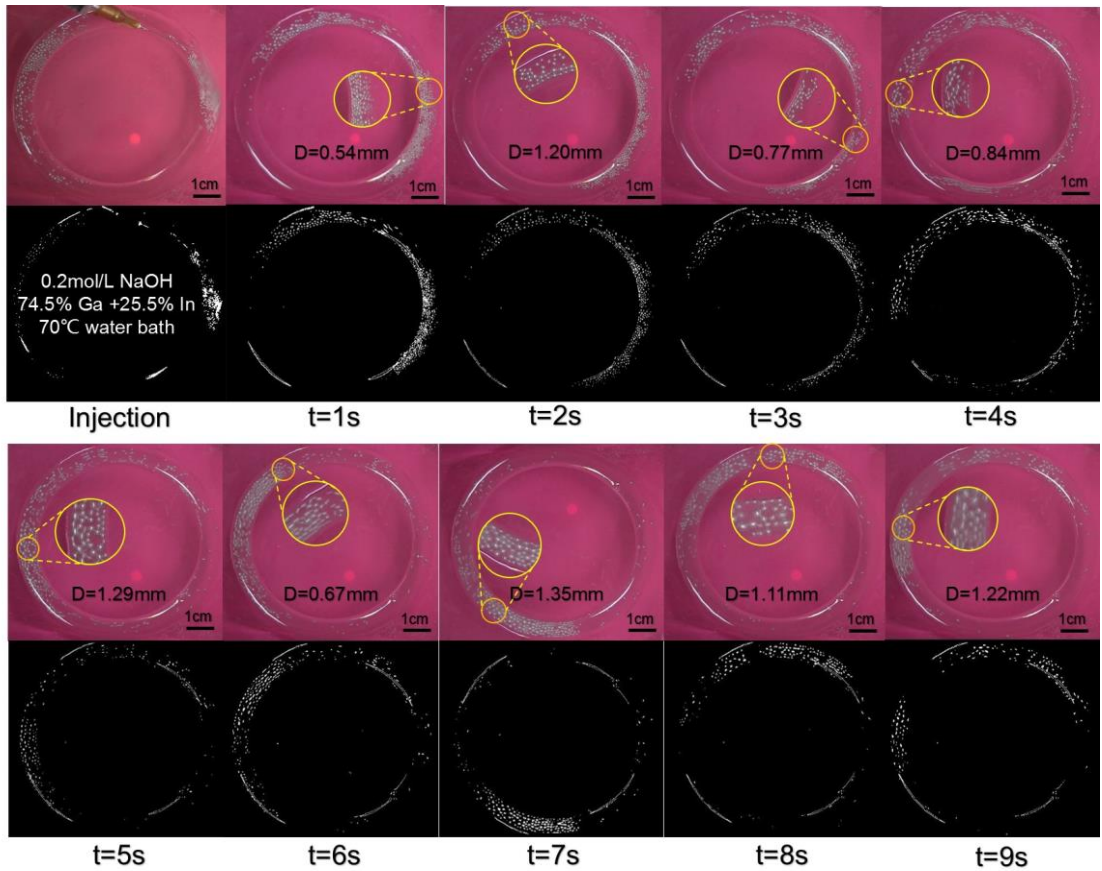

FIG. 4. Transient distributions of the running liquid metal motors in four quadrant regions I, II, III, IV. Here white pixel number in black and white images was used to estimate the motors distribution.

From repetitive measurements, we can find that motors in the circle are neither stable in channels nor distributed uniformly around different positions. In most times motors aggregate together and travel in roughly the same direction. When the cluster arrived at a certain space of specific quadrant as shown in the image, there would appear a peak on the corresponding curve of the right hand side in the Fig. 4. The peak travels and transforms among different curves indicating the cluster movement from one quadrant to another. The cluster can also split into two smaller clusters or join as one big cluster with some individual motors moving freely away from the big one. Because the size of the motors in a cluster cannot be precisely kept as the same, they have diverse locomotion velocity (also moving direction) thus collision would happen occasionally. Overall, the initial running direction for a stream of liquid metal motors seems follow its original injection direction. But if two separate streams collide together, the subsequent directions for each of the streams may be disturbed. Clearly, further microscopic interpretations are needed in the near future.

The present principle for spontaneously generating liquid metal tiny motors has generality. To test whether the same phenomenon would also happen in other electrolytes, additional experiments were carried out in the 0.2mol/L NaCl solution. As is anticipated, the same phenomenon does occur (Fig. 5). There are no evident differences between NaCl solution and NaOH case regarding the basic motor behaviors. This is because, no matter what kind of electrolyte is used, galvanic cell reaction would happen. Due to different electrochemical activities between EGaIn and aluminum, galvanic cell reaction would happen among them and the surrounding electrolyte. Droplets are just propelled by the generated interior electricity inside the sphere, the induced flow nearby and the hydrogen bubbles.

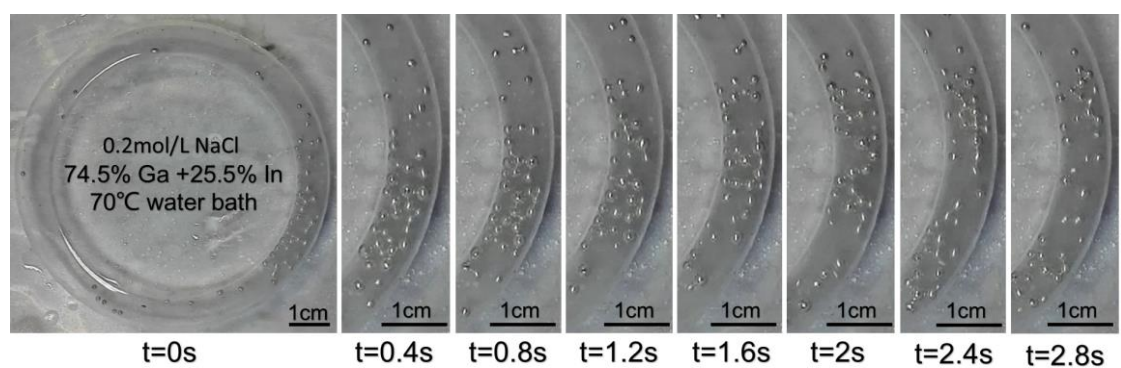

FIG. 5. Transient images for the running liquid metal tiny-motors in NaCl solution.

## 4. Discussion

It should be pointed out that, the present mechanisms for the self-powered tiny-motors are different from the existing ones where small-scale synthetic motors generally rely on converting the chemical energy into movement and forces. Impressive progress has been made over the past decade and a variety of microscale motors based on different design principles have been exhibited. Most motors are shaped like rods while propelled by micro bubbles generation. Gases can come from chemical reaction between metallic motor coat and acid solution or electrolyte catalysis reaction in which motors perform as catalyst [12].

In the present method, tiny motors are not only propelled by the bubbles but also driven by the electricity generated through the galvanic cell reaction. This mechanism is similar to former observation on electricity induced flow of liquid metal [16, 17]. A basic difference between the present work and the former is that all the previous liquid metal machines rely on the driving of external electrical device while the present one is an entirely self-fueled machine. Aluminum is adhered to liquid metal droplet and acts as chemical fuel in Al-water or Al-alkaline galvanic cell reaction. With electricity and bubble driving mechanisms working together, motors can have much higher motion efficiency.

Because of the varied electrochemical activity between EGaIn and Aluminum, the fuel could react with both NaOH and NaCl solution on the surface of liquid metal droplets. It is such reaction that breaks up symmetry of the electric double layer. When dispersing the liquid metal droplets inside the electrolyte, reaction between NaOH and Al proceeds as: $2Al+2NaOH+2H_2O=2NaAlO_2+3H_2\uparrow$ on the surface of sphere droplets. While in NaCl solution, reaction occurs as $2Al+6H_2O=2Al(OH)_3+3H_2$. For NaOH alkaline solution, anodic reaction is $6H_2O + 6e^- = 3H_2\uparrow + 6OH^-$ and cathodic reaction is $2Al + 8OH^- - 6e^- = 2AlO_2^- + 4H_2O$. Clearly, in both cases, it is such interiorly generated electricity and the gases there that drive the liquid metal droplet to run here and there.

According to Lippman's equation [17], one has $\gamma(V)=\gamma_0-cV^2/2$ (where $\gamma$, $c$, and $V$ are respectively the surface tension, the capacitance, and the potential difference across the electrical double layer of the liquid metal sphere, $\gamma_0$ is the maximum surface tension when V=0). Therefore, with the continuously generated electrical voltage from the galvanic cell, the surface tension of electric double layer of the liquid metal droplets will be varied. When such non-uniform electric field appears, the surface tension becomes asymmetric which would induce imbalance of the pressure difference *p* and thus motion of the motors. This is because there exists an approximate relation between $p$ and the surface tension, i.e., $p=2\gamma/r$, where, $1/r$ is the curvature of the droplet surface. In previous researches [16], it has been found that an external electric field would drive liquid metal droplets to transform between different morphologies and move in electrolyte. The present motor also follows the same rule. An only difference is that it is entirely based on the motor itself rather than the outside electrical field. This allows more freedom for the motors to be flexibly used.

**5. Conclusion**

In summary, we have found that a large quantity of self-powered liquid metal tiny-motors can be spontaneously fabricated via a rather straightforward way. Such motors swiftly running inside the electrolyte are enabled through fueling a tiny piece of aluminum to EGaIn alloy to convert chemical energy into propulsion power. Clearly, the prepared functional solutions can be stored without requesting extra requirement. Once needed, just injecting such mixtures to the desired electrolyte will make tiny motors. The motor amount, size and velocity etc. can all be well controlled through regulating the related factors. Overall, the basic characteristics and fundamental mechanisms disclosed would shed light for future works on complex fluid mechanics, droplet machines including injectable devices etc. It also opens opportunity for designing future soft-robot or microfluidic systems with high efficiency.

**Acknowledgments:**

This work is partially supported by the NSFC under Grant 81071225.

**References**

[1] T. Krupenkin, J. A. Taylor, Nature Comm. **2**, 1454 (2011).

[2] K. Q. Ma, J. Liu, J. Phys. D: Appl. Phy. **40**, 4722 (2007).

[3] S. Cheng, Z. G. Wu, Lab on a Chip **12**, 2782 (2012).

[4] M. Gao, L. Gui, Lab on a Chip **14**, 1866 (2014).

[5] Y. Zheng, Z. Z. He, J. Yang, J. Liu, Scientific Reports **4**, 4588 (2014).

[6] T. Hutter, W. A. C. Bauer, S. R. Elliott, W. T. S. Huck, Adv. Funct. Mater. **22**, 2624 (2012).

[7] Y. Yu, Q. Wang, L. T. Yi, J. Liu, Adv. Eng. Materials **16**, 255 (2014).

[8] W. Gao, J. Wang, Nanoscale **6**, 10486 (2014).

[9] W. Gao, M. D'Agostino, V. Garcia-Gradilla, J. Orozco, J. Wang, Small **9**, 467 (2013).

[10] F. Z. Mou, C. Chen, H. Ma, Y. Yin, Q. Wu, J. Guan, Angew. Chem. Int. Edit **52**, 7208 (2013).

[11] F. Z. Mou, C. R. Chen, Q. Zhong, Y. X. Yin, H. R. Ma, J. G. Guan, ACS Appl. Mater. Inter. **6**, 9897 (2014).

[12] G. Zhao, M. Viehrig, M. Pumera, Lab on a Chip **10**, 1930 (2013).

[13] W. Gao, R. Dong, S. Thamphiwatana, J. Li, W. Gao, L. Zhang, et al., ACS Nano **9**, 117 (2015).

[14] T. E. Mallouk, A. Sen, Scientific American pp.72-77 (2009).

[15] S. Y. Tang, V. Sivan, K. Khoshmanesh, A. P. O'Mullane, X. K. Tang, B. Gol, et al, Nanoscale **5**, 5949 (2013).

[16] L. Sheng, J. Zhang, and J. Liu, Adv. Mater. **26**, 6036 (2014).

[17] S. Y. Tang, K. Khoshmanesh, V. Sivan, P. Petersen, A. P. O'Mullane, D. Abbott, et al., PNAS **111**, 3304 (2014).

[18] Y. G. Deng and J. Liu, Appl. Phys. A **95**, 907 (2009).